\documentclass[twocolumn,aps,prl,final,superscriptaddress,showpacs]{revtex4-1}
\usepackage[latin9]{inputenc}
\usepackage{color}
\usepackage{amsmath}
\usepackage{graphicx}
\usepackage[unicode=true,pdfusetitle,
 bookmarks=true,bookmarksnumbered=true,bookmarksopen=true,bookmarksopenlevel=1,
 breaklinks=true,pdfborder={0 0 1},backref=false,colorlinks=true]
 {hyperref}
\hypersetup{
 linkcolor=blue,urlcolor=blue,citecolor=blue,pdfstartview={FitH},hyperfootnotes=false}

\makeatletter
\usepackage{dcolumn}
\usepackage{bm}

\usepackage{times}

\makeatother

\begin{document}
\title{Ultrahigh energy storage density in epitaxial AlN/ScN superlattices }
\author{Zhijun Jiang}
\affiliation{MOE Key Laboratory for Nonequilibrium Synthesis and Modulation of
Condensed Matter, School of Physics, Xi\textquoteright an Jiaotong
University, Xi\textquoteright an 710049, China}
\affiliation{Key Laboratory of Computational Physical Sciences (Ministry of Education),
State Key Laboratory of Surface Physics, and Department of Physics,
Fudan University, Shanghai 200433, China}
\affiliation{Physics Department and Institute for Nanoscience and Engineering,
University of Arkansas, Fayetteville, Arkansas 72701, USA }
\author{Bin Xu}
\email{binxu19@suda.edu.cn}

\affiliation{Jiangsu Key Laboratory of Thin Films, School of Physical Science and
Technology, Soochow University, Suzhou 215006, China}
\author{Hongjun Xiang}
\email{hxiang@fudan.edu.cn}

\affiliation{Key Laboratory of Computational Physical Sciences (Ministry of Education),
State Key Laboratory of Surface Physics, and Department of Physics,
Fudan University, Shanghai 200433, China}
\affiliation{Collaborative Innovation Center of Advanced Microstructures, Nanjing
210093, China}
\affiliation{Shanghai Qi Zhi Institute, Shanghai 200232, China}
\author{L. Bellaiche}
\affiliation{Physics Department and Institute for Nanoscience and Engineering,
University of Arkansas, Fayetteville, Arkansas 72701, USA }
\begin{abstract}
Dielectric and antiferroelectric materials are particularly promising
for high-power energy-storage applications. However, relatively low
energy density greatly hinders their usage in storage technologies.
Here, we report first-principles-based calculations predicting that
epitaxial and initially non-polar AlN/ScN superlattices can achieve
an ultrahigh energy density of up to 200 J/cm$^{\textrm{3}}$, accompanied
by an ideal efficiency of 100\%. We also show that high energy density
requires the system being neither too close nor too far from a ferroelectric
phase transition under zero electric field. A phenomenological model
is further proposed to rationalize such striking features. 
\end{abstract}
\maketitle
Electrostatic capacitors that are based on dielectric or antiferroelectric
materials are promising energy storage components in various electronic
applications because of their higher power density, faster charging/discharging
rates, and better stability when compared with supercapacitors and
batteries \cite{Chu2006,Hao2013,Patel2014,Li2015,Chauhan2015,Peng2015,Prateek2016,Yao2017,Xu2017,Huang2018,Yang2019,Pan2019,Zou2019,Li2020,Kim2020,Peddigari2019}.
However, their relatively low energy density is the main bottleneck
towards energy-storage applications, while intensive efforts have
been devoted to find novel compounds with improved storage properties.
In general, the recoverable energy density ($U_{\textrm{rec}}$) is
determined by the work done by the electric field during discharging,
$U_{\textrm{rec}}=\int_{P_{\textrm{r}}}^{P_{\textrm{max}}}EdP$, where
$P_{\textrm{max}}$ and $P_{\textrm{r}}$ are the maximum and remnant
polarization, respectively. To achieve high energy density, it is
usually required to have high $P_{\textrm{max}}$, low $P_{\textrm{r}}$,
and high breakdown strength \cite{Chauhan2015,Peng2015,Prateek2016,Yao2017,Xu2017,Huang2018}.
The efficiency is defined as $\eta=[U_{\textrm{rec}}/(U_{\textrm{rec}}+U_{\textrm{loss}})]\times100\%$,
where $U_{\textrm{loss}}$ is the energy loss density. Therefore,
a linear dielectric material has an energy density ($U$) of $U=\frac{1}{2}\varepsilon_{0}\varepsilon_{\mathrm{r}}E^{2}$,
where $\varepsilon_{0}$ is the vacuum permittivity, $\varepsilon_{\mathrm{r}}$
is the relative permittivity, and $E$ is the electric field. Obviously,
high values of permittivity $\varepsilon_{\mathrm{r}}$ and breakdown
strength $E_{\textrm{break}}$ lead to large energy density \cite{zhou2016,zhou2018,Li2018}.
On the other hand, nonlinear dielectrics -- usually including ferroelectrics,
antiferroelectric, and relaxors -- have either low energy density
(in ferroelectrics) or dissipative energy losses that can reduce efficiency
(in antiferroelectrics and relaxors). For instance, the energy density
is low for a ferroelectric material, due to large remnant polarization
\cite{Li2018}.

To maximize both energy density and efficiency, one can in fact imagine
a nonlinear type dielectric material that can combine the advantages
of a ferroelectric and a linear dielectric, i.e., possessing large
polarization under high field, and being nonpolar under zero field
as well as reversible. This would require the nonlinear dielectrics
to have large $\varepsilon_{\mathrm{r}}$ and high $E_{\textrm{break}}$
\cite{Yang2019,Pan2019,Zou2019,Li2020,Kim2020,Peddigari2019}. A promising
candidate is the III-V semiconductor-based systems made by mixing
AlN and ScN to form Al$_{1-x}$Sc$_{x}$N solid solutions or AlN/ScN
superlattices, that have been attracting much attention due to their
potential for high piezoelectric and electro-optic responses \cite{Akiyama2009,Akiyama2009-1,Tasnadi2010,Zhang2013,Daoust2017,Fichtner2019,Alam2019,Jiang2019,Yazawa2021,Yasuoka2020}.
Note that a recent experiment observed a ferroelectric switching in
Al$_{1-x}$Sc$_{x}$N films, with a remnant polarization reaching
a very large value -- in excess of 1.0 C/m$^{2}$ \cite{Fichtner2019}
-- which contrasts with the case of pure AlN that is of wurtzite-type
structure \cite{Wright1995,Bungaro2000} and is polar but not ferroelectric.
Note that for larger Sc compositions (above 43\%), Al$_{1-x}$Sc$_{x}$N
can be non-polar \cite{Akiyama2009,Akiyama2009-1,Fichtner2019}.

The aim of this Letter is to demonstrate and explain why very promising
energy-storage can be achieved in (Al,Sc)N systems being initially
in a non-polar phase but at close proximity with a ferroelectric state
that is easy to reach under electric fields.

Here, first-principles calculations are performed on (001) epitaxial
1$\times$1 AlN/ScN superlattices within the local density approximation
(LDA) to the density functional theory (DFT), as implemented in the
ABINIT package \cite{Gonze2002}. The epitaxial strain is defined
as $\eta_{\textrm{in}}=(a-a_{\textrm{eq}})/a_{\textrm{eq}}$, where
$a_{\textrm{eq}}$ is the in-plane lattice constant corresponding
to the equilibrium $P\bar{6}m2$ phase of bulk 1$\times$1 AlN/ScN
superlattices \cite{Jiang2019}. A 6 $\times$ 6 $\times$ 4\textbf{
}\textit{k}-point mesh and a plane-wave kinetic energy cutoff of 50
Hartree and of 60 Hartree are employed for structural relaxation.
The higher value of 60 Hartree is used to minimize the Pulay error
in the stress. A $dc$ electric field, $E$, is applied along the
$[001]$ direction ($c$-axis) by adopting the method developed in
Refs.~\cite{Nunes1994,Nunes2001,Souza2002,Zwanziger2012}. For each
field-induced structure, the electrical polarization, $P$, along
the $c$-axis is computed from the Berry phase method \cite{Smith1993,Resta1994}
and $P$-$E$ curves are then obtained. Note that, for each chosen
strain and applied $dc$ electric field, the in-plane lattice vectors
are kept fixed during the simulations while the out-of-plane lattice
vector and atomic positions are fully relaxed, until the force on
each atom is smaller than 10$^{-6}$ Hartree/Bohr -- in order to
simulate \textit{epitaxial} films under electric field.

Due to Landauer\textquoteright s paradox, the theoretical electric
field is typically larger compared with measurement \cite{Xu2017,Jiang2018,Lu2019,Chen2019,Jiang2020},
and the electric fields considered in this Letter have been rescaled
by a factor of 1/3, in order that the $P$-$E$ loop of the 1$\times$1
AlN/ScN system under a $-$3\% strain matches the experimental one
corresponding to a Al$_{0.57}$Sc$_{0.43}$N film \cite{Fichtner2019}
-- as demonstrated in Fig.~S1 of the Supplemental Material (SM)
\cite{SM}. Note that such rescaling of computational electric fields
is material-dependent but not phase-dependent for a given material
\cite{Jiang2018}. For instance, Ref. \cite{Jiang2018} demonstrated
that a rescaling of the electric field is able to reproduce very well
the electric field-\textit{versus}-temperature phase diagram of the
complex Pb(Mg,Nb)O$_{3}$ system, with this phase diagram involving
different phases (e.g., relaxor paraelectric \textit{versus} ferroelectric)
and even different orders (first \textit{versus} second) of the phase
transitions. Regarding the magnitude of the (renormalized) fields
to be considered here, it is first worthwhile to indicate that a recent
experiment reported a feasible electric field as high as $\simeq$5
MV/cm in Al$_{1-x}$Sc$_{x}$N films \cite{Fichtner2019}. Moreover,
one can also estimate the intrinsic breakdown field of such systems
empirically, which depends on the band gap \cite{Wang2006}. Our predicted
gap of 3.1 eV after correction (to account for the typical underestimation
of LDA) of the 1$\times$1 AlN/ScN superlattice ground state \cite{Jiang2019}
yields an intrinsic breakdown field, $E_{\textrm{break}}$, of 6.3
MV/cm, which will be the maximum fields considered here. Note that
the estimated breakdown field is reasonable since (1) Yazawa \textit{et
al}. reported the highest applied electric field ($\sim$6 MV/cm)
in epitaxial Al$_{0.7}$Sc$_{0.3}$N film \cite{Yazawa2021}; and
(2) Yasuoka \textit{et al}. also reported very large coercive fields
of the order of 4-7 MV/cm and maximum applicable electric fields of
the order 5-10 MV/cm in Al$_{1-x}$Sc$_{x}$N films \cite{Yasuoka2020}.
We also used another method \cite{Kim2016,Kim2016-1} to estimate
the intrinsic breakdown field, which provides a value of $\sim$6.59
MV/cm, that is rather close to the 6.3 MV/cm field used in our manuscript
(see the SM \cite{SM}).

\begin{figure}
\includegraphics[width=8cm]{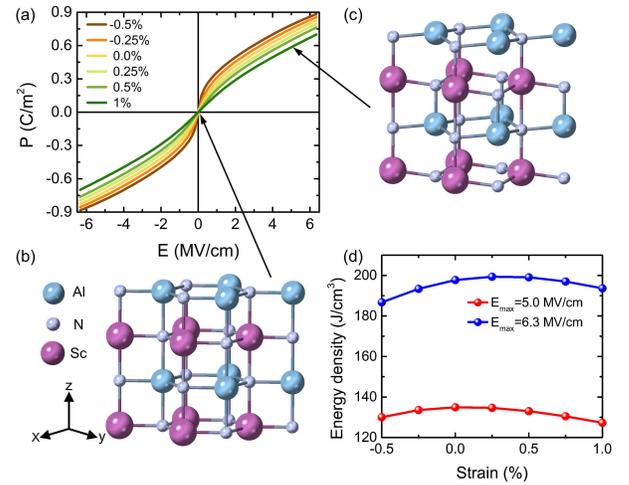}

\caption{(a) $P$-$E$ hysteresis curves with different epitaxial strains with
the ground state being a hexagonal-derived structure in 1$\times$1
AlN/ScN superlattices. Crystal structures of 1\% tensile strain under
electric fields of (b) 0 MV/cm and (c) 5 MV/cm, respectively, in 1$\times$1
AlN/ScN superlattices. (d) Energy density versus misfit strain for
two different ending electric fields of 5.0 MV/cm and 6.3 MV/cm. \label{fig:P-E loops}}
\end{figure}

For the 1$\times$1 AlN/ScN superlattices under different epitaxial
strains, previous studies have predicted the existence of different
phases \cite{Alam2019,Jiang2019}, including a wurtzite-derived structure
(polar $P3m1$ space group with a polarization along the $c$-axis)
and a hexagonal-derived structure (non-polar $P\bar{6}m2$ space group).
They are related to each other by a continuous change in the internal
parameter ($u$) and the axial ratio ($c/a$) \cite{Jiang2019}. Figure~\ref{fig:P-E loops}(a)
shows the $P$-$E$ curves for strain values larger than $-$0.5\%
(from compressive to tensile), with the hexagonal-derived structure
as ground state {[}see Fig.~\ref{fig:P-E loops}(b) for such structure
at 1\% strain under zero field{]}. The structure associated with the
electric field of 5 MV/cm is polar and close to the wurtzite-derived
structure {[}see Fig.~\ref{fig:P-E loops}(c){]}. 

When the strain is between $-$5\% and $-$1\% (as shown in Fig.~S1(b)
of the SM \cite{SM}), the $P$-$E$ loops show typical single hysteresis
of ferroelectrics with switching of polarization between the two degenerate
wurtzite-derived structures, and the magnitudes of the critical fields
depend very strongly on strain. Similar strong dependency was also
observed in Al$_{1-x}$Sc$_{x}$N solid solutions with respect to
the Sc composition \cite{Fichtner2019}; therefore, it suggests that
varying composition in the solid solution or changing the strain for
a fixed alloy or superlattice may involve similar physics. Due to
the strong strain dependency, the hysteresis rapidly shrinks with
reduced compressive strain, and completely vanishes beyond $-$0.5\%
strain, as depicted in Fig.~\ref{fig:P-E loops}(a). In this range
of strains, the ground state becomes the non-polar $P\bar{6}m2$ hexagonal-derived
phase. More interestingly, the $P$-$E$ curve is very nonlinear at
small compressive strain (e.g., $-$0.5\%) -- starting with large
dielectric susceptibility at small field and saturating at elevated
field -- while it becomes more linear as the strain increases (e.g.,
1\%). And the transition to the polar wurtzite-type phase is predicted
to be continuous and fully reversible upon increasing and decreasing
electric fields.

Let us now check the energy storage properties for the 1$\times$1
AlN/ScN superlattices for misfit strains varying between $-$0.5\%
to $+$1\%. The first feature to realize is that the energy efficiency
is automatically 100\% because the charging and discharging processes
are completely reversible, as revealed by the continuous $P$-$E$
curves of Fig.~\ref{fig:P-E loops}(a). The energy density is depicted
in Fig.~\ref{fig:P-E loops}(d) for epitaxial strain ranging between
$-$0.5\% and $+$1\%, with the maximal applied electric field being
5 MV/cm that has been realized experimentally \cite{Fichtner2019}
in this type of nitride semiconductors, and 6.3 MV/cm estimated as
the intrinsic breakdown field. It is striking that the energy density
reaches (slightly strain-dependent) values varying between 127 and
135 J/cm$^{3}$ and between 187 and 200 J/cm$^{3}$ when using 5 MV/cm
and 6.3 MV/cm for the largest applied field, respectively. All these
energy densities and efficiencies are thus predicted to be giant,
since they are much larger than those measured in lead-based and lead-free
dielectric thin films \cite{Peng2015,Pan2019,Li2020,Kim2020,Park2014,Ma2015,Sun2016,Pan2018},
such as (Bi$_{0.5}$Na$_{0.5}$)$_{0.9118}$La$_{0.02}$Ba$_{0.0582}$(Ti$_{0.97}$Zr$_{0.03}$)O$_{3}$
(BNLBTZ) epitaxial relaxor (154 J/cm$^{\textrm{3}}$ and 97\% for
a maximum applied field, $E_{\textrm{max}}$ $=$ 3.5 MV/cm) \cite{Peng2015},
BiFeO$_{3}$-BaTiO$_{3}$-SrTiO$_{3}$ solid solutions (112 J/cm$^{\textrm{3}}$
and 80\% for $E_{\textrm{max}}$ $=$ 4.9 MV/cm) \cite{Pan2019},
Pb$_{0.88}$Ca$_{0.12}$ZrO$_{3}$ (91.3 J/cm$^{\textrm{3}}$ and
85.3\% for $E_{\textrm{max}}$ $=$ 5 MV/cm) \cite{Li2020}, and ion-bombarded
relaxor ferroelectric 0.68Pb(Mg$_{1/3}$Nb$_{2/3}$)O$_{3}$-0.32PbTiO$_{3}$
(PMN-PT) films (133 J/cm$^{\textrm{3}}$ and 75\% for $E_{\textrm{max}}$
$=$ 5.9 MV/cm) \cite{Kim2020}. The 100\% efficiency and the giant
energy densities suggest that the 1$\times$1 AlN/ScN superlattices
are highly promising for energy-storage applications.

\begin{figure}
\includegraphics[width=8cm]{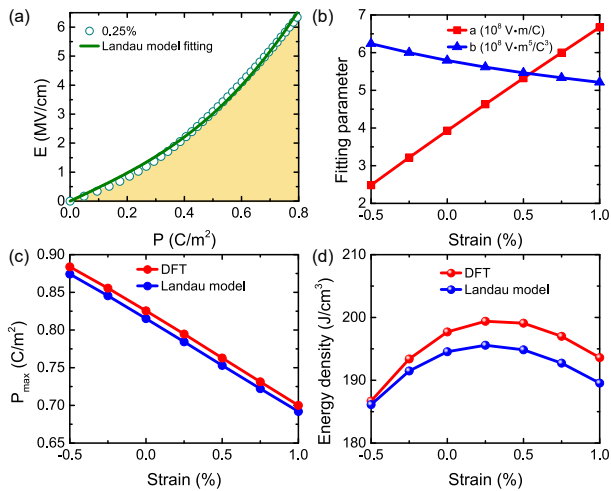}

\caption{(a) Electric field versus polarization for strains at 0.25\% (the
solid green line represents the fit of the DFT results by the Landau
model), with the yellow area showing the energy density. (b) Evolution
with strain of fitting parameters $a$ and $b$ (see text). Strain-dependence
of (c) $P_{\textrm{max}}$ and (d) energy density results obtained
from DFT and Landau model. These data correspond to a maximal applied
electric field of $E_{\textrm{break}}$ $=$ 6.3 MV/cm. \label{fig:Landau model}}
\end{figure}

To understand these results, let us recall that, a nonlinear dielectric
with no energy dissipation has an energy density $U$ has the following
form:

\begin{equation}
U=\int_{0}^{P_{\textrm{max}}}EdP,\label{eq:energy density}
\end{equation}
where $P_{\textrm{max}}$ is the maximum polarization, exhibited at
the highest applied field, $E_{\textrm{max}}$. Figure~\ref{fig:Landau model}(a)
displays the (positive) electric field applied along the $[001]$
pseudo-cubic direction as a function of polarization for one particular
misfit strain at 0.25\%. The area shown in yellow of Fig.~\ref{fig:Landau model}(a)
is the energy density.

The energy density in Fig.~\ref{fig:P-E loops}(d) can be understood
based on a simple Landau-type free energy that describes the behavior
of a nonlinear dielectric:

\begin{equation}
F=\frac{1}{2}aP^{2}+\frac{1}{4}bP^{4}-EP,\label{eq:Landau}
\end{equation}
where $a$ and $b$ are coefficients of the quadratic and quartic
terms, respectively, and they are positively defined for a paraelectric
or non-polar phase.

Noting that at equilibrium condition, we have $\frac{\partial F}{\partial P}=0$,
which leads to

\begin{equation}
E=aP+bP^{3}.\label{eq:Landau-1}
\end{equation}

Thus, the polarization is zero without the application of an electric
field. Consequently, the energy density can be written in the simple
following form:

\begin{equation}
U=\int_{0}^{P_{\textrm{max}}}(aP+bP^{3})dP=\frac{1}{2}aP_{\textrm{max}}^{2}+\frac{1}{4}bP_{\textrm{max}}^{4}.\label{eq:energy density Landau}
\end{equation}

As shown in Fig.~\ref{fig:Landau model}(a), the electric field versus
polarization ($E$-$P$) data for strains at 0.25\% can be very well
fitted by Eq.~(\ref{eq:Landau-1}). In fact, Eq.~(\ref{eq:Landau-1})
fits remarkably well the DFT $E$-$P$ data for all the considered
strain ranging between $-$0.5\% and $+$1\% (see Fig.~S3 of the
SM \cite{SM}). Figure~\ref{fig:Landau model}(b) shows the fitting
parameters $a$ and $b$ as a function of epitaxial strain ranging
between $-$0.5\% and $+$1\%, for electric fields up to the intrinsic
breakdown field $E_{\textrm{max}}$ = $E_{\textrm{break}}$ = 6.3
MV/cm. Interestingly, Fig.~\ref{fig:Landau model}(b) reveals that
the fitting parameter $a$ linearly increases when the strain increases
from $-$0.5\% to $+$1\% ($a$ varies from 2.5$\times$10$^{8}$
V$\cdot$m/C to 6.7$\times$10$^{8}$ V$\cdot$m/C). In contrast,
the $b$ parameter linearly decreases with strain, with the change
of $b$ being much smaller in percentage than $a$ and taking values
ranging between 6.2$\times$10$^{8}$ V$\cdot$m$^{5}$/C$^{3}$ and
5.2$\times$10$^{8}$ V$\cdot$m$^{5}$/C$^{3}$. Note that the $a$
parameter of the Landau expansion is negative in our phenomenology
model when the ground state of the superlattice is ferroelectric (see
Fig.~S4 of the SM \cite{SM}).

Figure~\ref{fig:Landau model}(c) displays the value of the polarization,
$P_{\textrm{max}}$, for $E_{\textrm{max}}$ = 6.3 MV/cm, as a function
of strain both from DFT calculations and the Landau model using the
DFT-fitted $a$ and $b$ in Eq.~(\ref{eq:Landau-1}). One can first
clearly see that, for any of these strains, the DFT and Landau model
provide nearly identical results. According to Fig.~\ref{fig:Landau model}(c),
$P_{\textrm{max}}$ is larger than 0.7 C/m$^{\textrm{2}}$ for any
considered strain ranging between $-$0.5\% and $+$1\%, which reflects
that the structure at such high field is wurtzite-like. Note that,
for $E_{\textrm{max}}=E_{\textrm{break}}$, $P_{\textrm{max}}$ linearly
decreases with strain from 0.88 to 0.70 C/m$^{\textrm{2}}$, which
can be simply understood by the fact that increasing strains towards
and within the tensile regime is known to reduce the out-of-plane
polarization in ferroelectric systems \cite{Alam2019,Yang2012,Chen2015}.

To understand the results in Fig.~\ref{fig:P-E loops}(d), we use
Eq.~(\ref{eq:energy density Landau}) to obtained the energy density
for different strains at $E_{\textrm{max}}$ $=$ 6.3 MV/cm. Figure~\ref{fig:Landau model}(d)
shows that the Landau model agrees rather well with the DFT-obtained
energy density, with an excellent agreement at small strain and a
slightly increasing discrepancy at larger strains (less than 2\% at
1.0\% strain), implying the validity of the Landau model. Equation~(\ref{eq:energy density Landau})
tells us that the energy density is the sum of two terms -- that
are $\frac{1}{2}aP_{\textrm{max}}^{2}$ and $\frac{1}{4}bP_{\textrm{max}}^{4}$.
The first contribution depends on the product of the $a$ parameter
and $P_{\textrm{max}}^{2}$, and is numerically found to increase
when the strain is enhanced between $-$0.5\% and $+$1\% (see Fig.~S5
of the SM \cite{SM}). The second contribution is the product between
the $b$ parameter and $P_{\textrm{max}}^{4}$, and decreases with
strain since $b$ slightly decreases when changing the strain from
$-$0.5\% to $+$1\% while $P_{\textrm{max}}$ is reduced with strain,
as a consequence of the aforementioned coupling between out-of-plane
polarization and in-plane strains {[}see Figs.~\ref{fig:Landau model}(b)
and \ref{fig:Landau model}(c){]}. Practically, the contribution of
$\frac{1}{2}aP_{\textrm{max}}^{2}$ (respectively, of $\frac{1}{4}bP_{\textrm{max}}^{4}$)
to the total energy density is 51\% (respectively, 49\%), 73\% (respectively,
27\%), and 84\% (respectively, 16\%) for strains at $-$0.5\%, 0.25\%,
and 1\%, respectively. Note also that the strain dependency of $\frac{1}{2}aP_{\textrm{max}}^{2}$
and $\frac{1}{4}bP_{\textrm{max}}^{4}$, along with the dependency
of $P_{\textrm{max}}$ on $E_{\textrm{max}}$ {[}see Fig.~\ref{fig:P-E loops}(a){]},
leads to a maximum value of the total energy density $U$ occurring
at a specific strain for a given $E_{\textrm{max}}$, while different
$E_{\textrm{max}}$ can further quantitatively modulate the performance.
For instance, such maximum occurs at 0\% and 0.25\% misfit strain
when the largest applied field is 5 MV/cm and 6.3 MV/cm, respectively
{[}see Fig.~\ref{fig:P-E loops}(d){]}.

As evidenced above, we numerically found that the energy density at
its maximum value, for our selected $E_{\textrm{max}}$ of 6.3 MV/cm,
mostly (i.e., at 73\%) depends on the $a$ parameter and $P_{\textrm{max}}$
via $\frac{1}{2}aP_{\textrm{max}}^{2}$. Let us thus fix $b$ as a
constant and vary the $a$ parameter for the case of $E_{\textrm{max}}$
$=$ 6.3 MV/cm, and compute the resulting $P_{\textrm{max}}$ and
energy density using Eqs.~(\ref{eq:Landau-1}) and (\ref{eq:energy density Landau}).
The results are shown in Fig.~S6 of the SM \cite{SM}), which reveal
that the energy density varies non-monotonically with the $a$ parameter
at a selected strain of 0.25\%, presenting a maximum, $U_{\textrm{max, b-fixed}}$,
for an intermediate $a$ value. Note that small $a$ gives smaller
density than $U_{\textrm{max, b-fixed}}$ because $\frac{1}{2}aP_{\textrm{max}}^{2}$
is reduced via the $a$ parameter; while larger $a$ also provides
an energy density being smaller than $U_{\textrm{max, b-fixed}}$
because now $P_{\textrm{max}}$ is weakened in $\frac{1}{2}aP_{\textrm{max}}^{2}$.
These facts automatically imply that there is a trade-off to have
large energy density: the system should not be too close (i.e., should
not have a very small $a$) but also not be too far (i.e., should
not possess a too large $a$) from a ferroelectric phase transition.

In summary, we have performed first-principles calculations to investigate
the energy-storage properties and out-of-plane strains in epitaxial
1$\times$1 AlN/ScN superlattices. We find that these systems exhibit
very high energy-densities up to 200 J/cm$^{3}$ and an ideal 100\%
efficiency. Interestingly, we also found that such systems can simultaneously
exhibit ultrahigh strain levels of about 10\%, as detailed in the
SM \cite{SM}. Note that InN can be used as a substrate to get 0.3\%
of strain \textcolor{black}{\cite{Jiang2019} } to achieve the predicted
performance, which is promising for energy storage in our proposed
system. Such unusual and promising features originate from high values
of the electric field that can be applied in these systems, and in
particular from a nonlinear electric-field-induced strong polarizability
of the non-polar hexagonal-like polymorph. A simple Landau model is
developed to analyze and understand these results, which also suggests
that (Al,Sc)N disordered solid solutions should also possess these
\textquotedblleft wunderbar\textquotedblright{} properties for compositions
lying in the non-polar regime near but not too close to the border
with the wurtzite-like structure. The proposed new concept of using
nonlinear and non-polar systems close to a ferroelectric state as
an energy storage material is general and we expect that other materials
with similar characteristics are also good candidates for energy storage. 
\begin{acknowledgments}
This work is supported by the National Natural Science Foundation
of China (Grants No.\ 11804138, No.\ 11825403, and No.\ 11991061),
Shandong Provincial Natural Science Foundation (Grant No.\ ZR2019QA008),
China Postdoctoral Science Foundation (Grants No.\ 2020T130120 and
No.\ 2018M641905), and \textquotedblleft Young Talent Support Plan\textquotedblright{}
of Xi\textquoteright an Jiaotong University (Grant No.\ WL6J004).
B.X.\ acknowledges financial support from National Natural Science
Foundation of China under Grant No.\ 12074277 and Natural Science
Foundation of Jiangsu Province (BK20201404), the startup fund from
Soochow University, and the support from Priority Academic Program
Development (PAPD) of Jiangsu Higher Education Institutions. L.B.\ acknowledges
the DARPA grant HR0011-15-2-0038 (MATRIX program). The Arkansas High
Performance Computing Center (AHPCC) of University of Arkansas and
HPC Platform of Xi'an Jiaotong University are also acknowledged. 
\end{acknowledgments}

\end{document}